\documentclass[a4paper, 12pt]{article}

\usepackage{amsmath, amsthm, amssymb}
\usepackage[top=2cm,nohead,left=2.7cm,right=2.7cm]{geometry}
\usepackage{setspace}
\usepackage{natbib}
\usepackage{hyperref}
\usepackage{tikz}
\usepackage{tikz-cd}
\usepackage{booktabs}

\newtheorem{theorem}{Theorem}
\newtheorem{lemma}{Lemma}
\newtheorem{proposition}{Proposition}
\newtheorem{corollary}{Corollary}
\newtheorem{definition}{Definition}

\title{Reachable Stability Lattices after Population Shocks}
\author{Yi-You Yang\thanks{Department of Applied Mathematics, Chung Yuan Christian University, Taoyuan City, Taiwan. E-mail address: yyyang@cycu.edu.tw}}
\date{}

\begin{document}

\maketitle

\begin{abstract}
This paper studies how one-sided population shocks, such as worker exits and
firm entries in senior-level labor markets, transform stability lattices in
many-to-many matching markets with contracts and substitutable preferences.
Deferred acceptance induces a re-equilibration map from the pre-shock to the
post-shock stability lattice. We show that this map preserves order and joins,
maps the pre-shock worker-optimal stable allocation to the greatest reachable
element, and maps the pre-shock firm-optimal stable allocation to the
post-shock firm-optimal stable allocation. Consequently, the reachable
post-shock set forms a finite lattice. The construction factors through the
firm-quasi-stable lattice, where the deferred-acceptance outcome coincides
with the limit of a monotone Tarski-type operator. Sequential shock invariance
shows that the reachable lattice is determined by the initial market and the
aggregate shock, not by the order of exits and entries.
\end{abstract}

\section{Introduction}
\label{sec:introduction}

Stable allocations in matching markets are often not unique. Under
substitutable preferences, however, this multiplicity is structured: in
matching markets with contracts, the set of stable allocations forms a lattice
under the Blair order
\citep{Blair1988,Alkan2002,Hatfield2005a,Hatfield2017}. This lattice orders
stable allocations by the welfare of one side of the market, identifies
side-optimal stable allocations, and defines joins and meets as extremal
stable allocations subject to these welfare comparisons. Thus, the stable set
is not merely a collection of equilibria; it has an order structure with
economic content.

This paper asks how that structure changes after a population shock. Many
matching markets are not fixed. In senior-level labor markets, for example, a
retirement, a resignation, or the creation of a new position may destabilize
an initially stable assignment and trigger a vacancy-chain adjustment. Once
the initial market and the aggregate shock are fixed, the relevant object is
not the full post-shock stability lattice. Some post-shock stable allocations
may be dynamically irrelevant from the pre-shock market. The central question
is which part of the post-shock stability lattice is reachable from pre-shock
stable allocations through re-equilibration, and whether this reachable part
inherits a lattice structure.

The model is a many-to-many matching market with contracts and substitutable
preferences. A one-sided population shock moves the market in the direction of
worker scarcity: workers may exit and firms may enter. The original market is
\((W,F')\), while the perturbed market is \((W',F)\), where
\(W'\subseteq W\) and \(F'\subseteq F\). Given a pre-shock stable allocation,
the natural projection keeps only the contracts involving surviving workers.
This projected allocation need not be stable in the perturbed market. It is,
however, firm-quasi-stable \citep{Sotomayor1996}. The lattice of
firm-quasi-stable allocations therefore provides the transitional state space
between the two stability lattices.

Re-equilibration from this transitional state is described by firm-proposing
deferred acceptance. We show that this process coincides with the limit of a
monotone Tarski-type operator on the firm-quasi-stable lattice. The resulting
map \(T^*\) is the order-theoretic projection from the transitional lattice to
the post-shock stability lattice: it sends each firm-quasi-stable allocation
to the least stable allocation that weakly Blair-dominates it. Composing
restriction with this projection gives the inter-market re-equilibration map
\[
\Phi:S(W,F')\to S(W',F).
\]

The main technical result is a compatibility theorem. It states that
population projection and re-equilibration commute: applying
re-equilibration before projection gives the same post-shock stable allocation
as projecting first and then re-equilibrating in the perturbed market. This
commutative relation is the mechanism behind the main structural results. It
implies that \(\Phi\) preserves joins. It also implies that the pre-shock
firm-optimal stable allocation is mapped to the post-shock firm-optimal stable
allocation, while the pre-shock worker-optimal stable allocation is mapped to
the greatest stable allocation reachable after the shock.

The image of \(\Phi\), denoted by \(\mathcal I(W,F';W',F)\), is the reachable
post-shock set. Since \(\Phi\) preserves joins, this set is join-closed inside
the post-shock stability lattice. Since it contains the least element of the post-shock stability lattice, it is bounded below within that lattice. 
These facts imply that
\(\mathcal I(W,F';W',F)\), ordered by the inherited worker Blair order, is a
finite lattice. Thus, the relevant post-shock object is not merely a collection
of outcomes. It is the lattice of post-shock stable allocations attainable
from pre-shock stable allocations under the given shock and re-equilibration
process.

The compatibility theorem also yields sequential shock invariance. If the same
one-sided population shock is decomposed through intermediate markets, direct
re-equilibration from the initial market to the final market gives the same
map as re-equilibration through the intermediate markets. Hence, the reachable
post-shock lattice depends on the initial market and the aggregate shock, but
not on the chronological decomposition of worker exits and firm entries.

Under the law of aggregate demand, the relationship between the transitional
and stable lattices becomes sharper. The post-shock stability lattice is a
sublattice of the firm-quasi-stable lattice. Moreover, joining the projected
pre-shock allocation with the post-shock firm-optimal stable allocation yields
a stable allocation. Together with the least-stable-upper-bound
characterization of re-equilibration, this gives
\[
\Phi(Y)=\Pi(Y)\vee_{W'}Y^F,
\]
where \(\Pi(Y)\) is the inherited allocation of surviving workers and \(Y^F\)
is the firm-optimal stable allocation of the perturbed market. This formula
yields a sharp prediction for entering firms: each entering firm receives
exactly its assignment at the post-shock firm-optimal stable allocation,
independently of the pre-shock stable allocation from which re-equilibration
starts.

The paper contributes to several strands of the matching literature. First, it
extends the lattice theory of stable allocations from fixed markets to
shock-induced comparisons across markets. Classical results describe the
internal lattice structure of stable allocations in a fixed market
\citep{Blair1988,Alkan2002,Hatfield2005a}. This paper studies the
inter-market transformation induced by a population shock and identifies the
reachable post-shock lattice associated with that shock.

Second, the paper builds on the literature on vacancy chains and
restabilization after market disruptions. The closest predecessor is
\citet{Blum1997}, who study senior-level labor markets in a one-to-one model.
They show that retirements and new positions transform stable matchings into
firm-quasi-stable matchings and that deferred acceptance restores stability
from such transitional states. They also establish monotonicity and invariance
properties of the re-equilibration process. Subsequent work extends
restabilization to many-to-one settings and studies related adjustment
procedures \citep{Cantala2004,Cantala2011}. These papers characterize how
stability is restored after disruptions. The present paper takes the induced
inter-market map itself as the main object and studies its order-theoretic
structure.

Third, the paper relates to operator, deferred-acceptance, and
quasi-stability approaches to matching. Tarski-type arguments and monotone
operators have been used to prove existence, derive lattice structure, and
study comparative statics in matching models
\citep{Adachi2000,Fleiner2003,Echenique2004,Hatfield2005a,Ostrovsky2008,Hatfield2017}.
More closely related are studies of quasi-stability and order-theoretic
restoration of stability \citep{Wu2018,Bonifacio2022,Bonifacio2024,Bonifacio2026}. These papers use Tarski operators on lattices of envy-free, worker-quasi-stable, or
firm-quasi-stable matchings to describe re-equilibration dynamics and to
characterize stable matchings as fixed points. Some of them also use such
operators to compute lattice operations within a fixed market. The present
paper builds on this operator approach but uses it for an inter-market
purpose. It treats the operator limit as a projection from the
firm-quasi-stable lattice to the stable lattice, proves that this projection
preserves joins, and shows that it coincides with firm-proposing deferred
acceptance. These properties allow the firm-quasi-stable lattice to serve not
only as a restoration domain within a fixed market, but also as the
transitional domain through which stability lattices are compared across
markets.

The remainder of the paper is organized as follows.
Section~\ref{sec:model} introduces the model, one-sided population shocks,
projection, and firm-quasi-stability. Section~\ref{sec:transitional} studies
the transitional lattice and the re-equilibration projection, and proves its
equivalence with firm-proposing deferred acceptance.
Section~\ref{sec:crossmarket} constructs the inter-market re-equilibration
map, proves the compatibility theorem, and derives join preservation,
boundary results, the reachable post-shock lattice, polarity, welfare
comparisons, and sequential shock invariance. Section~\ref{sec:lad} derives
the explicit formula and entrant-firm prediction under the law of aggregate
demand. Section~\ref{sec:conclusion} concludes. Proofs omitted from the main
text are collected in the Appendix.

\section{Model and Population Shocks}
\label{sec:model}

This section introduces the standard model of many-to-many matching markets with contracts and the class of population shocks studied in the paper. 
\subsection{Matching with Contracts}

A \emph{matching market with contracts} consists of a finite set of workers \(W\), a finite set of firms \(F\), and a finite set of contracts \(X\).
Each contract \(x\in X\) is associated with a worker \(w(x)\in W\) and a firm \(f(x)\in F\).

Assume throughout the paper that $W'\subseteq W$ and $F'\subseteq F$.
Let
\[
X_{W',F'}:=\{x\in X:w(x)\in W' \text{ and } f(x)\in F'\}
\]
denote the set of contracts available in the \emph{submarket} \((W',F')\). 
When only the worker side is restricted, we write
\(
X_{W'}:=X_{W',F}.
\)
When only the firm side is restricted, we write
\(
X_{F'}:=X_{W,F'}.
\)

An \emph{allocation} is a set of contracts \(Y\subseteq X\). For an agent \(i\in W\cup F\), let
\[
Y_i:=\{x\in Y:i\in\{w(x),f(x)\}\}
\]
denote the set of contracts in \(Y\) involving agent \(i\). 
For $A\subseteq W\cup F$, denote
\[
Y_A:=\bigcup_{i\in A}Y_i.
\]
We also write
\[
w(Y):=\{w\in W:Y_w\neq \emptyset\},
\qquad
f(Y):=\{f\in F:Y_f\neq \emptyset\}.
\]

Each agent \(i\in W\cup F\) has a strict preference relation \(\succ_i\) over the collection of subsets of \(X_i\). 
Given two allocations \(Y\) and \(Y'\),  we write 
\(
Y\succeq_i Y'
\) 
if
\(
Y_i\succeq_i Y'_i.
\)
The induced choice function of agent $i$ is defined by 
\[
C_i(Y):=\max_{\succeq_i}\{Z:Z\subseteq Y_i\}\quad\text{for}\quad Y\subseteq X,
\]
and the rejection function is thus defined by
\[
R_i(Y):=Y_i\setminus C_i(Y)\quad\text{for}\quad Y\subseteq X.
\]
For a set of agents \(A\subseteq W\cup F\),  the aggregate choice and rejection functions are defined by
\[
C_A(Y):=\bigcup_{i\in A}C_i(Y)\quad\text{and}\quad R_A(Y):=\bigcup_{i\in A}R_i(Y)\quad\text{for}\quad Y\subseteq X.
\]

\begin{definition}
The preferences of agent \(i\in W\cup F\) are \emph{substitutable} if, for all
\(Y\subseteq Z\subseteq X\),
\[
C_i(Z)\cap Y\subseteq C_i(Y).
\]
\end{definition}
Substitutability is equivalent to \emph{monotonicity} of the rejection function: for all \(Y\subseteq Z\subseteq X\),
\[
R_i(Y)\subseteq R_i(Z).
\]

\begin{definition}
The choice function \(C_i\) is \emph{path independent} if, for all
\(Y,Z\subseteq X\),
\[
C_i(Y\cup Z)=C_i(C_i(Y)\cup Z).
\]
\end{definition}

Because choice functions are induced by preferences, they satisfy  \emph{irrelevance of rejected contracts} (IRC):
\[
C_i(Y)=C_i(Z)\quad\text{whenever}\quad C_i(Z)\subseteq Y\subseteq Z\subseteq X.
\]
Thus, substitutability is equivalent to path independence in our model \citep{Aizerman1981}. 

Throughout the paper, all agents' preferences are assumed to be substitutable.
The same property applies to the aggregate choice functions \(C_W\), \(C_F\), \(C_{W'}\), and \(C_{F'}\).

\subsection{Blair Orders and the Stability Lattice}

For any allocation $Y$, the set of contracts acceptable to workers in $W'$ relative to $Y$ is defined by 
\[
\Gamma_{W'}(Y):=\{x\in X:x\in C_{W'}(Y\cup\{x\})\}.
\]
An allocation \(Y\) is \emph{individually rational} if
\(
C_{W}(Y)=C_{F}(Y)=Y.
\)
For any individually rational allocation $Y\subseteq X_{W',F'}$, we have $Y\subseteq \Gamma_{W'}(Y)$.

\begin{definition}
An allocation \(Y\subseteq X_{W',F'}\) is \emph{blocked} in market \((W',F')\) if there
exists a nonempty set of contracts \(Z\subseteq X_{W',F'}\setminus Y\) such that
\[
Z\subseteq C_{W'}(Y\cup Z)\cap C_{F'}(Y\cup Z).
\]
An allocation \(Y\subseteq X_{W',F'}\) is \emph{stable} in market \((W',F')\) if it is individually rational and is not blocked in $(W',F')$.
\end{definition}

Under substitutability, blocking by a set of contracts is equivalent to blocking by a single contract. 
Thus, an individually rational allocation $Y\subseteq X_{W',F'}$ is stable in $(W',F')$ if and only if there is no contract \(x\in X_{W',F'}\setminus Y\) such that
\[
x\in C_{W'}(Y\cup\{x\})\cap C_{F'}(Y\cup\{x\}).
\]
Let
\(
S(W',F')
\)
denote the set of stable allocations in market \((W',F')\).

The following characterization will be used throughout the paper.
The symmetric characterization, with the roles of workers and firms
interchanged, will also be used when convenient.

\begin{proposition}
\label{prop:stability}
An allocation \(Y\subseteq X_{W',F'}\) is stable in market \((W',F')\) if and only if
\[
Y=C_{F'}(\Gamma_{W'}(Y)).
\]
\end{proposition}

The set of individually rational allocations is ordered by the preferences of either side of the market $(W',F')$. 
For two individually rational allocations \(Y\) and \(Y'\), we say that \(Y\) \emph{weakly Blair-dominates} \(Y'\) for workers in $W'$, and write
\(
Y\succeq_{W'}^B Y',
\)
if
\[
C_{W'}(Y\cup Y')=Y_{W'}.
\]
Similarly, \(Y\) \emph{weakly Blair-dominates} \(Y'\) for firms in $F'$, written
\(
Y\succeq_{F'}^B Y',
\)
if
\[
C_{F'}(Y\cup Y')=Y_{F'}.
\]

When the relevant aggregate choice function is path independent, the corresponding Blair dominance relation is a partial order over the set of individually rational allocations. 
The classical lattice theorem states that stable allocations in a fixed market form a complete lattice under the Blair order \citep{Blair1988}.

\begin{theorem}
\label{thm:fixed-market-lattice}
The set \(S(W',F')\) is a nonempty complete lattice under \(\succeq_{W'}^B\).
\end{theorem}

For \(Y,Y'\in S(W',F')\), denote their \emph{join} and \emph{meet} within the stability lattice
by
\[
Y\tilde{\vee}_{W'} Y'
\quad\text{and}\quad
Y\tilde{\wedge}_{W'} Y',
\]
respectively. 
Thus, \(Y\tilde{\vee}_{W'} Y'\) is the least stable allocation that weakly Blair-dominates both \(Y\) and \(Y'\) for workers in $W'$, while \(Y\tilde{\wedge}_{W'} Y'\) is the greatest stable allocation that is weakly Blair-dominated by both \(Y\) and \(Y'\).

Since \(S(W',F')\) is a complete lattice under \(\succeq_{W'}^B\), define
\[
Y^{W'}(W',F')
:=
\max_{\succeq_{W'}^B} S(W',F')\quad\text{and}\quad
Y^{F'}(W',F')
:=
\min_{\succeq_{W'}^B} S(W',F').
\]

We call \(Y^{W'}(W',F')\) the \emph{worker-optimal} stable
allocation and \(Y^{F'}(W',F')\) the \emph{worker-pessimal} stable allocation. 
The notation anticipates the firm-side interpretation established in Theorem~\ref{thm:polarity}. 
When the market is clear from context, we write
these allocations simply as \(Y^{W'}\) and \(Y^{F'}\).

\subsection{One-Sided Population Shocks}

This paper studies shocks that transform an original market \((W,F')\) into a perturbed market \((W',F)\).
Thus, some existing workers may leave the market, while new firms may enter. 
We refer to such a change as a \emph{one-sided population shock} or a \emph{worker-scarcity shock}. 

For an allocation \(Y\subseteq X_{F'}\) in the original market, define its restriction to the surviving workers by the associated projection:
\[
\Pi(Y):=Y_{W'}.
\]
When \(Y\) is a stable allocation in market \((W,F')\), the projected allocation \(\Pi(Y)\) is not necessarily a stable allocation in the post-shock market \((W',F)\). 
The next definition identifies the transitional state space into which \(\Pi(Y)\) falls.

\subsection{Firm-Quasi-Stability}

Firm-quasi-stability is a relaxation of stability \citep{Sotomayor1996}. 
It requires that every contract currently held by firms still be chosen by firms from the set of contracts acceptable to those workers.

\begin{definition}
An allocation \(Y\subseteq X_{W'}\) is \emph{firm-quasi-stable} in market
\((W',F)\) if
\[
Y\subseteq C_F(\Gamma_{W'}(Y)).
\]
Let
\(
Q(W',F)
\)
denote the set of firm-quasi-stable allocations in market \((W',F)\).
Analogously, an allocation \(Y \subseteq X_{F'}\) is firm-quasi-stable in market \((W, F')\) if 
\[
Y \subseteq C_{F'}(\Gamma_{W}(Y)).
\]
Let \(Q(W, F')\) denote the set of such allocations.
\end{definition}

The empty allocation is firm-quasi-stable, so $Q(W',F)$ is nonempty.
Firm-quasi-stability implies individual rationality under substitutable preferences.
Proposition~\ref{prop:stability} ensures that every stable allocation is firm-quasi-stable.
The following proposition shows that firm-quasi-stability is preserved by the projection map $\Pi$.

\begin{proposition}
\label{prop:projection-to-quasi-stability}
For \(Y\in Q(W,F')\),  \(\Pi(Y)\in Q(W',F)\).
\end{proposition}

The next section studies the structure of this transitional domain and shows
that \(Q(W',F)\) is itself a complete lattice. Re-equilibration will then be described as a projection from \(Q(W',F)\) onto the post-shock stability lattice
\(S(W',F)\).

\section{Transitional Lattices and Re-equilibration}
\label{sec:transitional}

This section studies the structure of this transitional domain \(Q(W',F)\) and the re-equilibration process on it. 
The main result is that \(Q(W',F)\) is a complete lattice and that re-equilibration is a join-preserving projection from \(Q(W',F)\) onto the post-shock stability lattice \(S(W',F)\).
This projection admits two equivalent interpretations: as the limit of a monotone Tarski operator and as the outcome of firm-proposing deferred acceptance.

\subsection{The Transitional Lattice}

We first establish the lattice structure of \(Q(W',F)\) under the worker Blair order \(\succeq_{W'}^B\).

The following elementary property will be used repeatedly. Under the worker Blair order, the set of contracts acceptable to workers is antitone.

\begin{lemma}
\label{lem:gamma-antitone}
Assume that \(Y,Y'\subseteq X_{W'}\) are individually rational allocations. 
If
\(
Y\succeq_{W'}^B Y'
\),
then
\(
\Gamma_{W'}(Y)\subseteq \Gamma_{W'}(Y')
\).
\end{lemma}

\begin{theorem}
\label{thm:transitional-lattice}
The set \(Q(W',F)\) is a nonempty complete lattice under
\(\succeq_{W'}^B\). For any \(Y,Y'\in Q(W',F)\), the join is
\[
Y\vee_{W'}Y'=C_{W'}(Y\cup Y'),
\]
and the meet is
\[
Y\wedge_{W'}Y'
=
C_{W'}\!\left(
\bigcup
\left\{
Z\in Q(W',F):
Y\succeq_{W'}^B Z
\text{ and }
Y'\succeq_{W'}^B Z
\right\}
\right).
\]
\end{theorem}

We write the join on \(Q(W,F')\) as \(\vee_W\). The projection induced by a
worker-scarcity shock is compatible with joins.

\begin{lemma}
\label{lem:projection-join-preserving}
The projection map \(\Pi:Q(W,F')\to Q(W',F)\) is join-preserving:
\[
\Pi(Y\vee_WY')=\Pi(Y)\vee_{W'}\Pi(Y')
\qquad
\text{for all }Y,Y'\in Q(W,F').
\]
Consequently, \(\Pi\) is order-preserving.
\end{lemma}

\subsection{The Re-equilibration Operator}

The transitional lattice \(Q(W',F)\) is the post-shock state space. Define, for
\(Y\in Q(W',F)\),
\[
T(Y):=C_{W'}(C_F(\Gamma_{W'}(Y))).
\]
This operator selects workers' choices from the contracts firms choose among
those acceptable to workers at \(Y\).

\begin{proposition}
\label{prop:T-properties}
For every \(Y\in Q(W',F)\):
\begin{enumerate}
\item[(i)] \(T(Y)\succeq_{W'}^B Y\);
\item[(ii)] \(T(Y)\in Q(W',F)\);
\item[(iii)] \(T(Y)=Y\) if and only if \(Y\in S(W',F)\).
\end{enumerate}
\end{proposition}

The next lemma records the monotonicity of the operator $T$.

\begin{lemma}
\label{lem:T-isotone}
If \(Y\succeq_{W'}^B Y'\) in \(Q(W',F)\), then
\(T(Y)\succeq_{W'}^B T(Y')\).
\end{lemma}

\subsection{The Projection onto Stability}

Since \(T\) is extensive and monotone on the finite lattice \(Q(W',F)\), the
sequence \[Y,T(Y),T^2(Y),\ldots\] reaches a fixed point for every
\(Y\in Q(W',F)\) in finite steps. 
Define \(T^*(Y)\) to be this limit.

\begin{theorem}
\label{thm:Tstar-limit}
For every \(Y\in Q(W',F)\), \(T^*(Y)\) is the least stable allocation that
weakly Blair-dominates \(Y\).
\end{theorem}

\begin{proof}
By Proposition~\ref{prop:T-properties}, the sequence \(\{T^t(Y)\}_{t\ge0}\)
is weakly increasing in \(Q(W',F)\) and hence, by finiteness, reaches a fixed
point. The fixed point is stable by Proposition~\ref{prop:T-properties}.

Let \(Z\in S(W',F)\) with \(Z\succeq_{W'}^B Y\). Since \(T(Z)=Z\), monotonicity
of \(T\) gives \(Z=T^t(Z)\succeq_{W'}^B T^t(Y)\) for every \(t\). At the limit,
\(Z\succeq_{W'}^B T^*(Y)\). Thus, \(T^*(Y)\) is the least stable allocation
weakly Blair-dominating \(Y\).
\end{proof}

Thus \(T^*:Q(W',F)\to S(W',F)\) is the order-theoretic projection from the
transitional lattice onto the stability lattice.

\begin{theorem}
\label{thm:Tstar-homomorphism}
The map \(T^*:Q(W',F)\to S(W',F)\) is order-preserving and join-preserving:
for all \(Y,Y'\in Q(W',F)\),
\[
Y\succeq_{W'}^B Y'
\quad\Longrightarrow\quad
T^*(Y)\succeq_{W'}^B T^*(Y'),
\]
and
\[
T^*(Y\vee_{W'}Y')
=
T^*(Y)\tilde{\vee}_{W'}T^*(Y'),
\]
where \(\tilde{\vee}_{W'}\) denotes the join in \(S(W',F)\).
\end{theorem}

\begin{proof}
Order preservation follows from Lemma~\ref{lem:T-isotone} by iteration.

For join preservation, \(Y\vee_{W'}Y'\) weakly Blair-dominates both \(Y\) and
\(Y'\). Hence, \(T^*(Y\vee_{W'}Y')\) is a stable upper bound of
\(T^*(Y)\) and \(T^*(Y')\), so
\(T^*(Y\vee_{W'}Y')\succeq_{W'}^B
T^*(Y)\tilde{\vee}_{W'}T^*(Y')\).

Conversely, \(T^*(Y)\tilde{\vee}_{W'}T^*(Y')\) is stable and weakly
Blair-dominates both \(Y\) and \(Y'\), and therefore weakly Blair-dominates
\(Y\vee_{W'}Y'\). By Theorem~\ref{thm:Tstar-limit}, it weakly Blair-dominates
\(T^*(Y\vee_{W'}Y')\). The two inequalities give the equality.
\end{proof}

The preceding results imply that the stability lattice inherits meets from the
transitional lattice, while joins are obtained by applying \(T^*\) to transitional
joins.

\begin{theorem}
\label{thm:stable-sublattice-operations}
For \(Y,Y'\in S(W',F)\),
\[
Y \tilde{\vee}_{W'} Y'=T^*(Y\vee_{W'}Y')
\quad\text{and}\quad
Y\tilde{\wedge}_{W'} Y' = Y\wedge_{W'} Y',
\]
where \(\tilde{\vee}_{W'}\) and \(\tilde{\wedge}_{W'}\) denote the lattice
operations in \(S(W',F)\).
\end{theorem}

\begin{proof}
Since \(T^*\) fixes stable allocations, join preservation gives
\[
T^*(Y\vee_{W'}Y')
=
T^*(Y)\tilde{\vee}_{W'}T^*(Y')
=
Y\tilde{\vee}_{W'}Y'.
\]

Let \(M=Y\wedge_{W'}Y'\). Since \(T^*\) is extensive, \(T^*(M)\succeq_{W'}^B M\).
Also, \(Y\) and \(Y'\) are stable upper bounds of \(M\). Hence, by
Theorem~\ref{thm:Tstar-limit}, both \(Y\) and \(Y'\) dominate \(T^*(M)\). Thus,
\(T^*(M)\) is a stable lower bound of \(Y\) and \(Y'\), so
\(Y\tilde{\wedge}_{W'}Y'\succeq_{W'}^B T^*(M)\). On the other hand,
\(M\) dominates every lower bound of \(Y\) and \(Y'\) in \(Q(W',F)\), and hence,
\(M\succeq_{W'}^B Y\tilde{\wedge}_{W'}Y'\). 
Therefore,
\[
T^*(M)\succeq_{W'}^B M\succeq_{W'}^B
Y\tilde{\wedge}_{W'}Y'\succeq_{W'}^B T^*(M),
\]
so \(M=Y\tilde{\wedge}_{W'}Y'\).
\end{proof}
 
\subsection{Deferred Acceptance as Decentralized Re-equilibration}

The projection \(T^*\) has a decentralized implementation.

\paragraph{Firm-proposing DA from \(Y^0\).}
Fix \(Y^0\in Q(W',F)\) and set \(X^0=X_{W'}\). At step \(t\ge1\), if
\(C_F(X^{t-1})\subseteq Y^{t-1}\), stop and return \(Y^{t-1}\). Otherwise,
choose \(x_t\in C_f(X^{t-1})\setminus Y^{t-1}\) for some firm \(f\), and set
\[
Y^t=C_{W'}(Y^{t-1}\cup\{x_t\}),
\qquad
X^t=X^{t-1}\setminus R_{W'}(Y^{t-1}\cup\{x_t\}).
\]
Since a contract is proposed at most once, the algorithm terminates in finitely
many steps.

\begin{proposition}
\label{prop:DA-properties}
Let \(\{(Y^t,X^t)\}_{t\ge0}\) be generated by DA in \((W',F)\). Then:
\begin{enumerate}
\item[(i)] \(Y^t\succeq_{W'}^B Y^{t-1}\) for every \(t\ge1\);
\item[(ii)] \(Y^t\in Q(W',F)\) and \(\Gamma_{W'}(Y^t)\subseteq X^t\) for every \(t\ge0\);
\item[(iii)] DA terminates at a stable allocation in \((W',F)\).
\end{enumerate}
\end{proposition}

The decentralized DA procedure implements the order-theoretic projection \(T^*\).

\begin{theorem}
\label{thm:DA-limit}
For every \(Y^0\in Q(W',F)\), every DA trajectory from \(Y^0\) terminates at
\(T^*(Y^0)\). Hence, the DA outcome is proposal-order invariant. We denote it by
\(\mathit{DA}(Y^0;W',F)\), and
\[
\mathit{DA}(Y^0;W',F)=T^*(Y^0).
\]
\end{theorem}

\begin{proof}
Let \(Y^{t^*}\) be the terminal allocation. By
Proposition~\ref{prop:DA-properties}, \(Y^{t^*}\in S(W',F)\) and
\(Y^{t^*}\succeq_{W'}^B Y^0\). Theorem~\ref{thm:Tstar-limit} therefore gives
\(Y^{t^*}\succeq_{W'}^B T^*(Y^0)\).

We prove the reverse inequality. Consider a proposal step \(t\), and set
\(A_t=C_F(\Gamma_{W'}(Y^{t-1}))\). Since \(Y^{t-1}\in Q(W',F)\),
\(Y^{t-1}\subseteq A_t\), and \(T(Y^{t-1})=C_{W'}(A_t)\). By path independence,
\[
C_{W'}(T(Y^{t-1})\cup Y^t)
=
C_{W'}(A_t\cup\{x_t\}).
\]

If \(x_t\notin Y^t\), then
\(x_t\in R_{W'}(Y^{t-1}\cup\{x_t\})\). Since
\(Y^{t-1}\subseteq A_t\), rejection monotonicity implies
\(x_t\in R_{W'}(A_t\cup\{x_t\})\). Hence, IRC gives
\[
C_{W'}(A_t\cup\{x_t\})=C_{W'}(A_t).
\]
If \(x_t\in Y^t\), then \(x_t\in\Gamma_{W'}(Y^{t-1})\). By
Proposition~\ref{prop:DA-properties}(ii),
\(\Gamma_{W'}(Y^{t-1})\subseteq X^{t-1}\). Since
\(x_t\in C_f(X^{t-1})\), substitutability implies \(x_t\in A_t\), and again
\[
C_{W'}(A_t\cup\{x_t\})=C_{W'}(A_t).
\]
Thus, \(C_{W'}(T(Y^{t-1})\cup Y^t)=T(Y^{t-1})\), so
\(T(Y^{t-1})\succeq_{W'}^B Y^t\).

By monotonicity of \(T\), this local inequality implies inductively that
\(T^t(Y^0)\succeq_{W'}^B Y^t\) for every proposal step \(t\). Since
\(T^*(Y^0)\) weakly Blair-dominates every iterate \(T^t(Y^0)\), we obtain
\(T^*(Y^0)\succeq_{W'}^B Y^{t^*}\). Antisymmetry yields
\(Y^{t^*}=T^*(Y^0)\).
\end{proof}
\begin{corollary}
\label{cor:restarting-invariance}
Let \(\{Y^t\}_{t=0}^{t^*}\) be a DA trajectory from \(Y^0\in Q(W',F)\). Then, for
every \(t=0,\ldots,t^*\),
\[
\mathit{DA}(Y^t;W',F)=\mathit{DA}(Y^0;W',F).
\]
\end{corollary}

This section has shown that, in the perturbed market \((W',F)\), the
transitional lattice \(Q(W',F)\) is projected onto the stability lattice
\(S(W',F)\) by \(T^*\), and that firm-proposing DA implements this projection.

For market comparisons, write \(T^*_{A,B}:Q(A,B)\to S(A,B)\) for the
re-equilibration projection in market \((A,B)\). Thus the map constructed above
is \(T^*_{W',F}\). The next section combines \(T^*_{W',F}\) with the restriction
\(\Pi\) induced by a worker-scarcity shock to obtain a canonical map between
pre- and post-shock stability lattices.

\section{The Inter-Market Re-equilibration Map}
\label{sec:crossmarket}

We now combine the restriction induced by the shock with the post-shock
re-equilibration projection. Since \(S(W,F')\subseteq Q(W,F')\),
Proposition~\ref{prop:projection-to-quasi-stability} implies that
\(\Pi(Y)\in Q(W',F)\) for every \(Y\in S(W,F')\). Hence,
\(T^*_{W',F}(\Pi(Y))\) is well defined.

\subsection{The Re-equilibration Map}

\begin{definition}
\label{def:reequilibration-map}
The \emph{inter-market re-equilibration map} induced by the shock from
\((W,F')\) to \((W',F)\) is
\[
\Phi:=T^*_{W',F}\circ \Pi:S(W,F')\to S(W',F).
\]
\end{definition}

Equivalently, \(\Phi\) is characterized by
\[
\begin{tikzcd}[column sep=3.2cm, row sep=large]
Q(W,F') 
\arrow[d, "\Pi"'] 
&
S(W,F') 
\arrow[l, hook', "\iota"'] 
\arrow[d, "\Phi"] 
\\
Q(W',F) 
\arrow[r, "T^*_{W',F}"'] 
&
S(W',F),
\end{tikzcd}
\]
where \(\iota\) denotes inclusion. By Theorem~\ref{thm:DA-limit},
\[
\Phi(Y)=\mathit{DA}(\Pi(Y);W',F)\quad\text{for}\quad Y\in S(W,F').
\]

\subsection{Compatibility of Projection and Re-equilibration}

The next result is the \emph{compatibility theorem} linking population projection and re-equilibration.

\begin{theorem}
\label{thm:commutative-diagram}
For every \(Y\in Q(W,F')\),
\[
T^*_{W',F}(\Pi(Y))
=
\Phi\!\left(T^*_{W,F'}(Y)\right).
\]
Equivalently, the diagram
\[
\begin{tikzcd}[column sep=3.2cm, row sep=large]
Q(W,F')
\arrow[d, "\Pi"']
\arrow[r, "T^*_{W,F'}"]
&
S(W,F')
\arrow[d, "\Phi"]
\\
Q(W',F)
\arrow[r, "T^*_{W',F}"']
&
S(W',F)
\end{tikzcd}
\]
commutes.
\end{theorem}

\begin{proof}
Let \(Y\in Q(W,F')\), and set \(Z=T^*_{W,F'}(Y)\). By
Theorem~\ref{thm:DA-limit}, \(Z\) is the outcome of some DA execution in
\((W,F')\) from \(Y\). Write this execution as
\(Y^0=Y,\ldots,Y^{t^*}=Z\), with available sets \(X^t\) and proposals \(x_t\).

For each \(t\), set \(\bar{Y}^t=\Pi(Y^t)\) and
\[
\bar{X}^t=(X^t\cap X_{W'})\cup X_{W',F\setminus F'}.
\]
Then \(\bar{X}^0=X_{W'}\). We show that, after deleting null steps, the
sequence \(\{\bar{Y}^t\}\) is a prefix of a DA execution in \((W',F)\).

If \(x_t\notin X_{W'}\), then \(x_t\) involves a worker outside \(W'\). Since
worker choice is componentwise, \(\bar{Y}^t=\bar{Y}^{t-1}\), and no contract in
\(X_{W'}\) is removed. Hence, this step is null after projection.

Suppose \(x_t\in X_{W'}\), and let \(f=f(x_t)\). Then \(f\in F'\) and
\(x_t\in C_f(X^{t-1})\setminus Y^{t-1}\). Since
\(\bar{X}^{t-1}_f=(X^{t-1}\cap X_{W'})_f\), substitutability gives
\(x_t\in C_f(\bar{X}^{t-1})\setminus \bar Y^{t-1}\). Thus \(x_t\) is an
admissible proposal in the perturbed market. Moreover, componentwise worker
choice gives
\[
\bar{Y}^t=C_{W'}(\bar{Y}^{t-1}\cup\{x_t\})
\]
and
\[
\bar{X}^t
=
\bar{X}^{t-1}\setminus R_{W'}(\bar{Y}^{t-1}\cup\{x_t\}).
\]
Thus the projected non-null step is exactly a DA step in \((W',F)\).

Consequently, the projected execution gives a DA prefix in \((W',F)\) from
\(\Pi(Y)\) to \(\Pi(Z)\). Extending this prefix to a full DA execution and using
restart invariance,
\[
\mathit{DA}(\Pi(Y);W',F)=\mathit{DA}(\Pi(Z);W',F).
\]
By Theorem~\ref{thm:DA-limit},
\[
T^*_{W',F}(\Pi(Y))=T^*_{W',F}(\Pi(Z)).
\]
Since \(Z=T^*_{W,F'}(Y)\) and
\(\Phi(Z)=T^*_{W',F}(\Pi(Z))\), the result follows.
\end{proof}

For \(Y\in S(W,F')\), Theorem~\ref{thm:commutative-diagram} reduces to the
definition of \(\Phi\). Its content is that the same commutative relation holds
on the whole transitional lattice \(Q(W,F')\).

\subsection{Join Preservation}

The compatibility theorem yields the main lattice implication of the
inter-market map.

\begin{theorem}
\label{thm:Phi-join-preserving}
The map \(\Phi:S(W,F')\to S(W',F)\) preserves joins: for all
\(Y,Y'\in S(W,F')\),
\[
\Phi(Y\tilde{\vee}_W Y')
=
\Phi(Y)\tilde{\vee}_{W'}\Phi(Y').
\]
\end{theorem}

\begin{proof}
For \(Y,Y'\in S(W,F')\), Theorem~\ref{thm:stable-sublattice-operations} gives
\(Y\tilde{\vee}_W Y'=T^*_{W,F'}(Y\vee_W Y')\). Hence, by
Theorem~\ref{thm:commutative-diagram},
\[
\Phi(Y\tilde{\vee}_W Y')
=
T^*_{W',F}\bigl(\Pi(Y\vee_W Y')\bigr)
=
T^*_{W',F}\bigl(\Pi(Y)\vee_{W'}\Pi(Y')\bigr),
\]
where the second equality uses Lemma~\ref{lem:projection-join-preserving}.
By Theorem~\ref{thm:Tstar-homomorphism}, the last term equals
\[
T^*_{W',F}(\Pi(Y))
\tilde{\vee}_{W'}
T^*_{W',F}(\Pi(Y'))
=
\Phi(Y)\tilde{\vee}_{W'}\Phi(Y').
\]
\end{proof}

Thus, a worker-scarcity shock induces a join-preserving transformation between the pre- and post-shock stability lattices. 
The standard order and meet comparisons follow.

\begin{corollary}
\label{cor:Phi-order-and-meet}
The map \(\Phi:S(W,F')\to S(W',F)\) is order-preserving. Moreover, for all
\(Y,Y'\in S(W,F')\),
\[
\Phi(Y)\tilde{\wedge}_{W'}\Phi(Y')
\succeq_{W'}^B
\Phi(Y\tilde{\wedge}_W Y').
\]
\end{corollary}

\subsection{Boundary Allocations and the Reachable Lattice}

The compatibility theorem also identifies the boundary points of the reachable
post-shock set. Define
\[
\mathcal I(W,F';W',F):=\Phi(S(W,F')).
\]
We call \(\mathcal I(W,F';W',F)\) the \emph{reachable post-shock set}.

\begin{corollary}[Lower boundary]
\label{cor:lower-boundary}
The re-equilibration map sends the worker-pessimal stable allocation of the
original market to the worker-pessimal stable allocation of the perturbed
market:
\[
\Phi\bigl(Y^{F'}(W,F')\bigr)=Y^F(W',F).
\]
In particular, \(Y^F(W',F)\in\mathcal I(W,F';W',F)\).
\end{corollary}

\begin{proof}
Apply Theorem~\ref{thm:commutative-diagram} to \(\emptyset\in Q(W,F')\).
Since \(\Pi(\emptyset)=\emptyset\),
\[
T^*_{W',F}(\emptyset)
=
\Phi\bigl(T^*_{W,F'}(\emptyset)\bigr).
\]
By Theorem~\ref{thm:Tstar-limit}, \(T^*_{W,F'}(\emptyset)=Y^{F'}(W,F')\)
and \(T^*_{W',F}(\emptyset)=Y^F(W',F)\). Hence,
\[
\Phi\bigl(Y^{F'}(W,F')\bigr)=Y^F(W',F).
\]
\end{proof}

\begin{corollary}[Upper boundary]
\label{cor:upper-boundary}
The worker-optimal stable allocation of the original market generates the
greatest reachable post-shock stable allocation:
\[
\Phi\bigl(Y^W(W,F')\bigr)
=
\max_{\succeq_{W'}^B}\mathcal I(W,F';W',F).
\]
\end{corollary}

\begin{proof}
Let \(Z\in\mathcal I(W,F';W',F)\). Then \(Z=\Phi(Y)\) for some
\(Y\in S(W,F')\). 
Since \(Y^W(W,F')\succeq_W^B Y\), Corollary~\ref{cor:Phi-order-and-meet}
implies
\[
\Phi\bigl(Y^W(W,F')\bigr)\succeq_{W'}^B Z.
\]
Thus \(\Phi(Y^W(W,F'))\) is the greatest element of
\(\mathcal I(W,F';W',F)\).
\end{proof}

\begin{theorem}
\label{thm:image-finite-lattice}
The reachable set \(\mathcal I(W,F';W',F)\), ordered by the worker Blair order
inherited from \(S(W',F)\), is a finite lattice. Its least element is
\(Y^F(W',F)\), its greatest element is \(\Phi(Y^W(W,F'))\), and its join is
inherited from \(S(W',F)\).
\end{theorem}

\begin{proof}
By Theorem~\ref{thm:Phi-join-preserving}, \(\mathcal I(W,F';W',F)\) is closed
under joins in \(S(W',F)\). Hence, the join of any two elements of
\(\mathcal I(W,F';W',F)\), computed in \(S(W',F)\), belongs again to
\(\mathcal I(W,F';W',F)\).

Corollary~\ref{cor:lower-boundary} shows that
\(Y^F(W',F)\in\mathcal I(W,F';W',F)\). Since \(Y^F(W',F)\) is the least element
of \(S(W',F)\), it is also the least element of \(\mathcal I(W,F';W',F)\).
Corollary~\ref{cor:upper-boundary} identifies
\(\Phi(Y^W(W,F'))\) as the greatest element of \(\mathcal I(W,F';W',F)\).

It remains to verify the existence of meets. Let
\(Z,Z'\in\mathcal I(W,F';W',F)\). The set of their common lower bounds in
\(\mathcal I(W,F';W',F)\) is nonempty, since it contains \(Y^F(W',F)\). Because
\(\mathcal I(W,F';W',F)\) is finite and join-closed, the join of these common
lower bounds, computed in \(S(W',F)\), belongs to \(\mathcal I(W,F';W',F)\) and
is their greatest common lower bound. Thus every pair in
\(\mathcal I(W,F';W',F)\) has both a join and a meet. Therefore
\(\mathcal I(W,F';W',F)\) is a finite lattice.
\end{proof}

Theorem~\ref{thm:image-finite-lattice} shows that the reachable stable allocations form a lattice inside \(S(W',F)\). 
Its least element is the post-shock worker-pessimal stable allocation; its greatest element is generated
by the pre-shock worker-optimal stable allocation. 
 
\subsection{Polarity and Welfare Comparison}

The preceding lattice results are stated in the worker Blair order. 
We next translate them into the corresponding firm-side comparison.

\begin{theorem} 
\label{thm:polarity}
Let \(Y\in S(W,F')\) and \(Y'\in S(W',F)\). Then 
\[ 
Y'\succeq^B_{W'} Y \quad\text{if and only if}\quad Y\succeq^B_{F'} Y'. 
\] 
\end{theorem}

When \(W'=W\) and \(F'=F\), Theorem~\ref{thm:polarity} reduces to the usual opposition-of-interests property. 
In particular, the worker-pessimal stable allocation is firm-optimal, and the worker-optimal stable allocation is firm-pessimal. 
Thus the least element identified in Theorem~\ref{thm:image-finite-lattice} is also the post-shock firm-optimal stable allocation.

\begin{corollary}
\label{cor:comparison}
For every \(Y\in S(W,F')\),
\[
\Phi(Y)\succeq^B_{W'}Y
\quad\text{and}\quad
Y\succeq^B_{F'}\Phi(Y).
\]
\end{corollary}

The welfare comparison is static: it compares the pre-shock allocation with the post-shock re-equilibrated allocation. 
We close the section with a dynamic invariance property, showing that the same final map is obtained when the shock
is decomposed into intermediate shocks.

\subsection{Sequential Shock Invariance}

A one-sided population shock may be decomposed into intermediate shocks without
changing the induced re-equilibration map.

\begin{theorem}
\label{thm:sequential-shock}
Assume that \(W'\subseteq \tilde W\subseteq W\) and
\(F'\subseteq \tilde F\subseteq F\). Let
\(\Phi_1:S(W,F')\to S(\tilde W,\tilde F)\) and
\(\Phi_2:S(\tilde W,\tilde F)\to S(W',F)\) be the re-equilibration maps induced
by the two shocks, and let \(\Phi:S(W,F')\to S(W',F)\) be the direct map. Then
\[
\Phi=\Phi_2\circ\Phi_1.
\]
\end{theorem}

\begin{proof}
Let \(\Pi_1(Y)=Y_{\tilde W}\), \(\Pi_2(Y)=Y_{W'}\), and \(\Pi(Y)=Y_{W'}\). Since
\(W'\subseteq \tilde W\), \(\Pi=\Pi_2\circ\Pi_1\). The upper square in
\[
\begin{tikzcd}[column sep=3.4cm, row sep=large]
Q(W,F')
\arrow[d, "\Pi_1"']
&
S(W,F')
\arrow[l, hook', "\iota"']
\arrow[d, "\Phi_1"]
\\
Q(\tilde W,\tilde F)
\arrow[d, "\Pi_2"']
\arrow[r, "T^*_{\tilde W,\tilde F}"']
&
S(\tilde W,\tilde F)
\arrow[d, "\Phi_2"]
\\
Q(W',F)
\arrow[r, "T^*_{W',F}"']
&
S(W',F)
\end{tikzcd}
\]
commutes by the definition of \(\Phi_1\), and the lower square commutes by
Theorem~\ref{thm:commutative-diagram}. Therefore, for every \(Y\in S(W,F')\),
\[
(\Phi_2\circ\Phi_1)(Y)
=
T^*_{W',F}\bigl((\Pi_2\circ\Pi_1)(Y)\bigr)
=
T^*_{W',F}(\Pi(Y))
=
\Phi(Y).
\]
Thus, \(\Phi=\Phi_2\circ\Phi_1\).
\end{proof} 

\section{Sharp Predictions under the Law of Aggregate Demand}
\label{sec:lad}

The preceding results require only substitutability. This section adds the law
of aggregate demand to obtain sharper structure. Under this additional
assumption, the stability lattice becomes a sublattice of the
firm-quasi-stable lattice, and the join of the projected pre-shock allocation
with the post-shock firm-optimal stable allocation yields the explicit
re-equilibration outcome.

\begin{definition}
\label{def:lad}
Agent \(i\)'s preferences satisfy the \emph{law of aggregate demand} if, for
all \(Z\subseteq Y\subseteq X\),
\[
|C_i(Z)|\le |C_i(Y)|.
\]
\end{definition}

\subsection{Sublattice Structure}

Under the law of aggregate demand, joining a firm-quasi-stable allocation with
a stable allocation already yields a stable allocation.

\begin{proposition}
\label{prop:lad-join-stability}
Under the law of aggregate demand, if \(Y\in Q(W',F)\) and \(Y'\in S(W',F)\), then
\[
Y\vee_{W'}Y'\in S(W',F).
\]
\end{proposition}

\begin{theorem}
\label{thm:lad-sublattice}
Under the law of aggregate demand, 
\(S(W',F)\) is a sublattice of 
\(Q(W',F)\).
Equivalently, for all \(Y,Y'\in S(W',F)\),
\[
Y\tilde{\vee}_{W'}Y'=Y\vee_{W'}Y'
\quad\text{and}\quad
Y\tilde{\wedge}_{W'}Y'=Y\wedge_{W'}Y'.
\]
\end{theorem}

\subsection{Explicit Representation of Re-equilibration}

Let
\[
Y^F:=Y^F(W',F)=T^*_{W',F}(\emptyset)
\]
be the worker-pessimal, equivalently firm-optimal, stable allocation of the
perturbed market.

\begin{theorem}
\label{thm:explicit-formula}
Under the law of aggregate demand,
for every \(Y\in S(W,F')\),
\[
\Phi(Y)=\Pi(Y)\vee_{W'}Y^F.
\]
\end{theorem}

\begin{proof}
By Proposition~\ref{prop:projection-to-quasi-stability},
\(\Pi(Y)\in Q(W',F)\). Since \(Y^F\in S(W',F)\),
Proposition~\ref{prop:lad-join-stability} implies that
\(\Pi(Y)\vee_{W'}Y^F\) is stable. It also weakly Blair-dominates \(\Pi(Y)\).

Let \(Z\in S(W',F)\) weakly Blair-dominate \(\Pi(Y)\). Since \(Y^F\) is the
least element of \(S(W',F)\), \(Z\succeq_{W'}^B Y^F\). Thus, \(Z\) is an upper
bound of \(\Pi(Y)\) and \(Y^F\) in \(Q(W',F)\), so
\(Z\succeq_{W'}^B \Pi(Y)\vee_{W'}Y^F\). Hence,
\(\Pi(Y)\vee_{W'}Y^F\) is the least stable allocation weakly Blair-dominating
\(\Pi(Y)\). By Theorem~\ref{thm:Tstar-limit},
\[
\Phi(Y)=T^*_{W',F}(\Pi(Y))=\Pi(Y)\vee_{W'}Y^F.
\]
\end{proof}

Thus, under the law of aggregate demand, post-shock re-equilibration has a
closed form: it is the worker-side join of the inherited allocation and the
post-shock firm-optimal stable allocation.

\subsection{Entering Firms}

The explicit formula gives a sharper prediction for firms that enter after the
shock. Let \(F^E:=F\setminus F'\).

\begin{corollary}
\label{cor:entrant-firms}
Under the law of aggregate demand,
for every \(Y\in S(W,F')\) and every entering firm \(f\in F^E\),
\[
\Phi(Y)_f=Y^F_f.
\]
\end{corollary}

Thus, each entering firm receives the same assignment as in the post-shock
firm-optimal stable allocation, independently of the pre-shock stable
allocation from which re-equilibration starts. 
 
\section{Concluding Remarks}
\label{sec:conclusion}

Stable allocations in matching markets are typically multiple, but this
multiplicity is structured: under substitutable preferences, stable allocations
form a lattice. This paper asks how that lattice structure changes after a
one-sided population shock.

The answer is not obtained by comparing the two stability lattices directly.
After the shock, the restriction of a pre-shock stable allocation need not be
stable in the post-shock market. It does, however, lie in the larger lattice of
firm-quasi-stable allocations. This transitional lattice provides the state
space on which post-shock re-equilibration is naturally defined.

The central construction is the inter-market re-equilibration map
\[
\Phi:S(W,F')\to S(W',F),
\qquad
\Phi(Y)=T^*_{W',F}(\Pi(Y)).
\]
It first restricts a pre-shock stable allocation to the surviving workers and
then projects the resulting transitional allocation back to stability in the
post-shock market. The compatibility theorem shows that restriction across
markets and re-equilibration within markets commute. This commutative structure
is the main technical mechanism behind the results.

The image of \(\Phi\) identifies the post-shock stable allocations that are
reachable from pre-shock stable allocations. This reachable set is not an
arbitrary subset of the post-shock stability lattice. It is a finite lattice
under the inherited worker Blair order. Its least element is the post-shock
firm-optimal stable allocation, while its greatest element is generated by the
pre-shock worker-optimal stable allocation. Thus the population shock induces a
structured transformation between stability lattices, rather than merely a
correspondence between individual stable allocations.

The same structure also disciplines the order in which shocks are implemented.
If the transition from \((W,F')\) to \((W',F)\) is decomposed through
intermediate markets, the resulting re-equilibration map is unchanged. In this
sense, the final reachable stable outcome is invariant to the sequential
decomposition of the one-sided population shock.

Under the law of aggregate demand, the relationship between the transitional and stable lattices becomes sharper. The stability lattice is then a sublattice of the firm-quasi-stable lattice, and the least-stable-upper-bound characterization of re-equilibration yields the explicit representation
\[
\Phi(Y)=\Pi(Y)\vee_{W'}Y^F.
\]
Hence, the post-shock outcome is determined by the inherited allocation of
surviving workers and the firm-optimal stable allocation of the perturbed
market. In particular, each entering firm receives the same assignment as in
the post-shock firm-optimal stable allocation, independently of the pre-shock
stable allocation.

Overall, the paper shows that although the full post-shock stability lattice
may contain stable allocations that are not dynamically reachable from the
pre-shock market, the reachable part remains a lattice and admits a canonical
description through projection and re-equilibration.

\appendix

\section{Proofs Omitted from the Main Text}
\label{app:proofs}

This appendix collects proofs of auxiliary results used in the main text. The
notation is the same as in the body of the paper.

\subsection{Proofs for Section~\ref{sec:model}}

\begin{proof}[Proof of Proposition~\ref{prop:stability}]
\((\Rightarrow)\)
Assume that \(Y\in S(W',F')\). 
Individual rationality implies
\(
Y\subseteq \Gamma_{W'}(Y)
\).
Let \(x\in C_{F'}(\Gamma_{W'}(Y))\). 
Since
\(x\in \Gamma_{W'}(Y)\) and \(Y\subseteq \Gamma_{W'}(Y)\), substitutability implies
\[
x\in C_{F'}(\Gamma_{W'}(Y))\cap (Y\cup\{x\})
\subseteq C_{F'}(Y\cup\{x\}).
\]
Also, \(x\in\Gamma_{W'}(Y)\) implies
\(
x\in C_{W'}(Y\cup\{x\})
\).
Stability therefore requires \(x\in Y\). 
Hence,
\[
C_{F'}(\Gamma_{W'}(Y))\subseteq Y.
\]
Together with \(Y\subseteq \Gamma_{W'}(Y)\), IRC and firm-side individual rationality imply
\[
C_{F'}(\Gamma_{W'}(Y))=C_{F'}(Y)=Y.
\]

\((\Leftarrow)\)
Assume that
\(
Y=C_{F'}(\Gamma_{W'}(Y))
\).
Then \(Y\subseteq \Gamma_{W'}(Y)\). 
Since
\[
C_{F'}(\Gamma_{W'}(Y))=Y\subseteq \Gamma_{W'}(Y),
\]
IRC implies
\[
C_{F'}(Y)=C_{F'}(\Gamma_{W'}(Y))=Y.
\]
Moreover, \(Y\subseteq \Gamma_{W'}(Y)\) implies \(Y\subseteq C_{W'}(Y)\), while
\(C_{W'}(Y)\subseteq Y\). Hence,
\[
C_{W'}(Y)=Y,
\]
so \(Y\) is individually rational.

Suppose, to the contrary, that there exists \(y\in X_{W',F'}\setminus Y\) such that
\[
y\in C_{W'}(Y\cup\{y\})\cap C_{F'}(Y\cup\{y\}).
\]
Then \(y\in \Gamma_{W'}(Y)\setminus Y\). Since
\[
C_{F'}(\Gamma_{W'}(Y))=Y\subseteq Y\cup\{y\}\subseteq \Gamma_{W'}(Y),
\]
IRC implies
\[
C_{F'}(Y\cup\{y\})=C_{F'}(\Gamma_{W'}(Y))=Y.
\]
This contradicts \(y\in C_{F'}(Y\cup\{y\})\). 
Hence, \(Y\) is stable in market \((W',F')\).
\end{proof}

\begin{proof}[Proof of Proposition~\ref{prop:projection-to-quasi-stability}]
Let \(Y\in Q(W,F')\). Then
\[
Y\subseteq C_{F'}(\Gamma_W(Y))\subseteq C_F(\Gamma_W(Y)).
\]

Take any \(x\in Y_{W'}\). Since \(x\in Y\subseteq C_{F'}(\Gamma_W(Y))\), we have
\(x\in \Gamma_W(Y)\). As \(w(x)\in W'\) and \(Y\) and \(Y_{W'}\) coincide on
workers in \(W'\), this implies \(x\in \Gamma_{W'}(Y_{W'})\). The same observation
gives \(\Gamma_{W'}(Y_{W'})\subseteq \Gamma_W(Y)\). Hence, by substitutability of
\(C_F\),
\[
x\in C_F(\Gamma_W(Y))\cap \Gamma_{W'}(Y_{W'})
\subseteq C_F(\Gamma_{W'}(Y_{W'})).
\]
Since \(x\in Y_{W'}\) was arbitrary, \(Y_{W'}\subseteq C_F(\Gamma_{W'}(Y_{W'}))\).
Thus \(\Pi(Y)=Y_{W'}\in Q(W',F)\).
\end{proof}

\subsection{Proofs for Section~\ref{sec:transitional}}

\begin{proof}[Proof of Lemma~\ref{lem:gamma-antitone}]
Let \(x\in \Gamma_{W'}(Y)\). Since \(Y\succeq_{W'}^B Y'\), 
\(C_{W'}(Y\cup Y')=Y\). By path independence,
\(C_{W'}(Y\cup Y'\cup\{x\})=C_{W'}(Y\cup\{x\})\). Hence,
\(x\in C_{W'}(Y\cup Y'\cup\{x\})\). Since
\(Y'\cup\{x\}\subseteq Y\cup Y'\cup\{x\}\), substitutability gives
\(x\in C_{W'}(Y'\cup\{x\})\). Thus \(x\in\Gamma_{W'}(Y')\).
\end{proof}

\begin{proof}[Proof of Theorem~\ref{thm:transitional-lattice}]
Let \(Y,Y'\in Q(W',F)\), and set \(Z^*=C_{W'}(Y\cup Y')\).
We first show that \(Z^*\in Q(W',F)\). Since \(Z^*=C_{W'}(Y\cup Y')\),
we have \(Z^*\subseteq \Gamma_{W'}(Z^*)\). If
\(x\in \Gamma_{W'}(Z^*)\), path independence gives
\[
x\in C_{W'}(Y\cup Y'\cup\{x\}).
\]
By substitutability,
\(x\in C_{W'}(Y\cup\{x\})\cap C_{W'}(Y'\cup\{x\})\), and hence,
\(\Gamma_{W'}(Z^*)\subseteq \Gamma_{W'}(Y)\cap \Gamma_{W'}(Y')\).

Now let \(z\in Z^*\). Since \(Z^*\subseteq Y\cup Y'\), either \(z\in Y\) or
\(z\in Y'\). If \(z\in Y\), then \(z\in C_F(\Gamma_{W'}(Y))\); if
\(z\in Y'\), then \(z\in C_F(\Gamma_{W'}(Y'))\). In either case, since
\(z\in Z^*\subseteq \Gamma_{W'}(Z^*)\) and
\(\Gamma_{W'}(Z^*)\subseteq \Gamma_{W'}(Y)\cap \Gamma_{W'}(Y')\),
substitutability of \(C_F\) implies
\(z\in C_F(\Gamma_{W'}(Z^*))\). Thus
\(Z^*\subseteq C_F(\Gamma_{W'}(Z^*))\), so \(Z^*\in Q(W',F)\).

It remains to verify that \(Z^*\) is the least upper bound. Path independence
implies \(C_{W'}(Z^*\cup Y)=Z^*\) and \(C_{W'}(Z^*\cup Y')=Z^*\), so
\(Z^*\succeq_{W'}^B Y,Y'\). If \(Z\in Q(W',F)\) satisfies
\(Z\succeq_{W'}^B Y,Y'\), then path independence gives
\[
C_{W'}(Z\cup Z^*)
=
C_{W'}(Z\cup Y\cup Y')
=
Z.
\]
Hence, \(Z\succeq_{W'}^B Z^*\), proving that \(Z^*=Y\vee_{W'}Y'\).

For the meet, let
\[
\mathcal L(Y,Y')
:=
\{Z\in Q(W',F):Y\succeq_{W'}^B Z
\text{ and }Y'\succeq_{W'}^B Z\}.
\]
This set is nonempty because \(\emptyset\in Q(W',F)\). Since \(Q(W',F)\) is
closed under \(\vee_{W'}\), the finite join
\[
Z_*:=\bigvee_{Z\in\mathcal L(Y,Y')} Z
=
C_{W'}\!\left(\bigcup_{Z\in\mathcal L(Y,Y')}Z\right)
\]
belongs to \(Q(W',F)\). By construction, \(Z_*\) dominates every common lower
bound. Moreover, since both \(Y\) and \(Y'\) dominate every element of
\(\mathcal L(Y,Y')\), they dominate their join \(Z_*\). Hence, \(Z_*\) is the
greatest lower bound of \(Y\) and \(Y'\).

Thus every pair in \(Q(W',F)\) has a join and a meet. Since \(X\) is finite,
\(Q(W',F)\) is a finite nonempty lattice, and therefore a complete lattice.
\end{proof}

\begin{proof}[Proof of Lemma~\ref{lem:projection-join-preserving}]
Let \(Y,Y'\in Q(W,F')\). Since aggregate worker choice is separable across
workers,
\[
\Pi(Y\vee_W Y')
=
\bigl(C_W(Y\cup Y')\bigr)_{W'}
=
C_{W'}(Y_{W'}\cup Y'_{W'})
=
\Pi(Y)\vee_{W'}\Pi(Y').
\]
Thus \(\Pi\) preserves joins.

If \(Y\succeq_W^B Y'\), then \(Y=Y\vee_WY'\). Hence,
\[
\Pi(Y)=\Pi(Y\vee_WY')=\Pi(Y)\vee_{W'}\Pi(Y'),
\]
so \(\Pi(Y)\succeq_{W'}^B\Pi(Y')\). Therefore, \(\Pi\) is order-preserving.
\end{proof}

\begin{proof}[Proof of Proposition \ref{prop:T-properties}]
Fix \(Y\in Q(W',F)\) and write \(A=C_F(\Gamma_{W'}(Y))\). Then
\(Y\subseteq A\) and \(T(Y)=C_{W'}(A)\). 
Since
\(T(Y)\subseteq T(Y)\cup Y\subseteq A\), IRC gives
\(C_{W'}(T(Y)\cup Y)=T(Y)\). 
Hence, \(T(Y)\succeq_{W'}^B Y\).

Next, \(T(Y)\) is individually rational: worker-side individual rationality follows from \(T(Y)=C_{W'}(A)\), while firm-side individual rationality follows from
\(T(Y)\subseteq A=C_F(\Gamma_{W'}(Y))\) and substitutability. 
By Lemma~\ref{lem:gamma-antitone}, \(T(Y)\succeq_{W'}^B Y\) implies
\(\Gamma_{W'}(T(Y))\subseteq \Gamma_{W'}(Y)\). 
Therefore,
\[
T(Y)\subseteq C_F(\Gamma_{W'}(Y))\cap \Gamma_{W'}(T(Y))
\subseteq C_F(\Gamma_{W'}(T(Y))),
\]
where the second inclusion follows from substitutability. 
Thus,
\(T(Y)\in Q(W',F)\).

If \(Y\in S(W',F)\), Proposition~\ref{prop:stability} gives
\(Y=C_F(\Gamma_{W'}(Y))\), and hence, \(T(Y)=C_{W'}(Y)=Y\). 
Conversely, suppose \(T(Y)=Y\).
Since \(Y\subseteq C_F(\Gamma_{W'}(Y))\), it suffices to prove the reverse inclusion. 
If \(x\in C_F(\Gamma_{W'}(Y))\setminus Y\), then
\(x\in\Gamma_{W'}(Y)\), while IRC applied to
\(Y=C_{W'}(C_F(\Gamma_{W'}(Y)))\) and
\(Y\subseteq Y\cup\{x\}\subseteq C_F(\Gamma_{W'}(Y))\) gives
\(C_{W'}(Y\cup\{x\})=Y\), a contradiction. 
Hence, \(Y=C_F(\Gamma_{W'}(Y))\), and Proposition~\ref{prop:stability} implies \(Y\in S(W',F)\).
\end{proof}

\begin{proof}[Proof of Lemma \ref{lem:T-isotone}]
Let \(Y\succeq_{W'}^B Y'\). Since firm-quasi-stability implies individual
rationality, Lemma~\ref{lem:gamma-antitone} gives
\(\Gamma_{W'}(Y)\subseteq \Gamma_{W'}(Y')\). Put
\(A=\Gamma_{W'}(Y)\), \(B=\Gamma_{W'}(Y')\), and
\(D=B\setminus R_F(A)\). By rejection monotonicity,
\(R_F(A)\subseteq R_F(B)\), and hence, \(C_F(B)\subseteq D\).

We claim that \(C_{W'}(D)=T(Y)\). Since \(Y\in Q(W',F)\),
\(Y\subseteq C_F(A)\subseteq D\). If \(x\in C_{W'}(D)\), then
\(Y\cup\{x\}\subseteq D\), so substitutability gives
\(x\in C_{W'}(Y\cup\{x\})\). Thus \(x\in A\). Since also \(x\in D\), we have
\(x\notin R_F(A)\), and hence, \(x\in C_F(A)\). Therefore
\(C_{W'}(D)\subseteq C_F(A)\subseteq D\), so IRC implies
\[
C_{W'}(D)=C_{W'}(C_F(A))=T(Y).
\]

Finally, \(C_F(B)\subseteq D\) implies \(T(Y')=C_{W'}(C_F(B))\subseteq D\).
Since \(T(Y)=C_{W'}(D)\) and
\(T(Y)\subseteq T(Y)\cup T(Y')\subseteq D\), IRC yields
\(C_{W'}(T(Y)\cup T(Y'))=T(Y)\). 
Hence,
\(T(Y)\succeq_{W'}^B T(Y')\).
\end{proof}

\begin{proof}[Proof of Proposition \ref{prop:DA-properties}]
We prove (i) and the invariant in (ii) by induction. The claim holds at \(t=0\).

Suppose it holds at \(t-1\). By path independence,
\(C_{W'}(Y^t\cup Y^{t-1})=Y^t\), so \(Y^t\succeq_{W'}^B Y^{t-1}\). Moreover,
if \(z\in\Gamma_{W'}(Y^t)\), then path independence gives
\(z\in C_{W'}(Y^{t-1}\cup\{x_t,z\})\). Substitutability therefore implies
\(z\in C_{W'}(Y^{t-1}\cup\{z\})\), and hence,
\(\Gamma_{W'}(Y^t)\subseteq \Gamma_{W'}(Y^{t-1})\subseteq X^{t-1}\).

No contract in \(\Gamma_{W'}(Y^t)\) is removed at step \(t\). Indeed, if
\(z\in\Gamma_{W'}(Y^t)\cap R_{W'}(Y^{t-1}\cup\{x_t\})\), then
\(C_{W'}(Y^{t-1}\cup\{x_t\})=Y^t\subseteq Y^t\cup\{z\}\subseteq
Y^{t-1}\cup\{x_t\}\). IRC implies
\(C_{W'}(Y^t\cup\{z\})=Y^t\), contradicting
\(z\in\Gamma_{W'}(Y^t)\). 
Thus, \(\Gamma_{W'}(Y^t)\subseteq X^t\).

It remains to show \(Y^t\in Q(W',F)\). If \(x_t\notin Y^t\), then IRC gives
\(Y^t=Y^{t-1}\), so the conclusion follows from the induction hypothesis.
Suppose \(x_t\in Y^t\). Then \(x_t\in\Gamma_{W'}(Y^{t-1})\). Since
\(x_t\in C_f(X^{t-1})\) and \(\Gamma_{W'}(Y^{t-1})\subseteq X^{t-1}\),
substitutability gives \(x_t\in C_F(\Gamma_{W'}(Y^{t-1}))\). Together with
\(Y^{t-1}\in Q(W',F)\), this implies
\(Y^t\subseteq C_F(\Gamma_{W'}(Y^{t-1}))\). Since
\(Y^t\subseteq\Gamma_{W'}(Y^t)\subseteq\Gamma_{W'}(Y^{t-1})\), substitutability
yields \(Y^t\subseteq C_F(\Gamma_{W'}(Y^t))\). 
Hence, \(Y^t\in Q(W',F)\).

Let \(t^*\) be the terminal step. Then
\(C_F(X^{t^*})\subseteq Y^{t^*}\subseteq
\Gamma_{W'}(Y^{t^*})\subseteq X^{t^*}\). IRC therefore gives
\(C_F(\Gamma_{W'}(Y^{t^*}))=C_F(Y^{t^*})=C_F(X^{t^*})\). Since
\(Y^{t^*}\in Q(W',F)\), it follows that
\(Y^{t^*}=C_F(\Gamma_{W'}(Y^{t^*}))\). By
Proposition~\ref{prop:stability}, \(Y^{t^*}\in S(W',F)\).
\end{proof}

\begin{proof}[Proof of Corollary~\ref{cor:restarting-invariance}]
Let \(\bar Y=\mathit{DA}(Y^0;W',F)=T^*(Y^0)\). By
Proposition~\ref{prop:DA-properties}, \(\bar Y\succeq_{W'}^B Y^t\succeq_{W'}^B
Y^0\). Since \(T^*(Y^t)\) is stable and weakly Blair-dominates \(Y^t\), it also
weakly Blair-dominates \(Y^0\). Theorem~\ref{thm:Tstar-limit} gives
\(T^*(Y^t)\succeq_{W'}^B \bar Y\). Conversely, \(\bar Y\) is stable and weakly
Blair-dominates \(Y^t\), so Theorem~\ref{thm:Tstar-limit} gives
\(\bar Y\succeq_{W'}^B T^*(Y^t)\). Hence, \(T^*(Y^t)=\bar Y\). The conclusion
follows from Theorem~\ref{thm:DA-limit}.
\end{proof} 
 
\subsection{Proofs for Section~\ref{sec:crossmarket}}

\begin{proof}[Proof of Corollary~\ref{cor:Phi-order-and-meet}]
If \(Y\succeq_W^B Y'\), then \(Y=Y\tilde{\vee}_W Y'\). Join preservation gives
\(\Phi(Y)=\Phi(Y)\tilde{\vee}_{W'}\Phi(Y')\), and hence,
\(\Phi(Y)\succeq_{W'}^B\Phi(Y')\).

Since \(Y\) and \(Y'\) both dominate \(Y\tilde{\wedge}_W Y'\), order
preservation implies that \(\Phi(Y\tilde{\wedge}_W Y')\) is a common lower
bound of \(\Phi(Y)\) and \(\Phi(Y')\). Therefore it is dominated by their meet,
which gives the stated inequality.
\end{proof}

\begin{proof}[Proof of Theorem~\ref{thm:polarity}]
Suppose first that \(Y'\succeq^B_{W'}Y\), so
\(C_{W'}(Y\cup Y')=Y'\). For every \(z\in Y'_{F'}\), substitutability gives
\(z\in C_{W'}(Y\cup\{z\})\), and hence, \(z\in\Gamma_W(Y)\). Since \(Y\) is
individually rational, \(Y\subseteq\Gamma_W(Y)\). Thus
\[
Y\cup Y'_{F'}\subseteq \Gamma_W(Y).
\]
By stability of \(Y\) in \((W,F')\),
\(C_{F'}(\Gamma_W(Y))=Y\). Since
\(Y\subseteq Y\cup Y'_{F'}\subseteq \Gamma_W(Y)\), IRC gives
\[
C_{F'}(Y\cup Y'_{F'})=Y.
\]
Because \(C_{F'}\) ignores contracts of firms outside \(F'\), this is
\(C_{F'}(Y\cup Y')=Y\). Hence, \(Y\succeq^B_{F'}Y'\).

The converse is symmetric. Suppose that \(Y\succeq^B_{F'}Y'\), so
\(C_{F'}(Y\cup Y')=Y\). Let
\[
\Gamma_F(Y'):=\{x\in X:x\in C_F(Y'\cup\{x\})\}.
\]
For every \(z\in Y_{W'}\), substitutability gives
\(z\in C_F(Y'\cup\{z\})\), and hence, \(z\in\Gamma_F(Y')\). Since \(Y'\) is
individually rational, \(Y'\subseteq\Gamma_F(Y')\). Thus
\[
Y'\cup Y_{W'}\subseteq \Gamma_F(Y').
\]
By the symmetric stability characterization for \(Y'\in S(W',F)\),
\(C_{W'}(\Gamma_F(Y'))=Y'\). Since
\(Y'\subseteq Y'\cup Y_{W'}\subseteq \Gamma_F(Y')\), IRC gives
\[
C_{W'}(Y'\cup Y_{W'})=Y'.
\]
Because \(C_{W'}\) ignores contracts of workers outside \(W'\), this is
\(C_{W'}(Y\cup Y')=Y'\). Hence, \(Y'\succeq^B_{W'}Y\).
\end{proof}

\begin{proof}[Proof of Corollary~\ref{cor:comparison}]
Since \(\Phi(Y)=T^*_{W',F}(\Pi(Y))\) and \(T^*_{W',F}\) is extensive,
\(\Phi(Y)\succeq^B_{W'}\Pi(Y)\). This is equivalent to
\(\Phi(Y)\succeq^B_{W'}Y\). The second assertion follows from
Theorem~\ref{thm:polarity}, applied to \(Y\in S(W,F')\) and
\(\Phi(Y)\in S(W',F)\).
\end{proof}

\subsection{Proofs for Section~\ref{sec:lad}}

\begin{proof}[Proof of Proposition~\ref{prop:lad-join-stability}]
Let \(Z=Y\vee_{W'}Y'=C_{W'}(Y\cup Y')\). Since \(Q(W',F)\) is closed under
joins, \(Z\in Q(W',F)\). Also \(Z\succeq_{W'}^B Y'\), so
Lemma~\ref{lem:gamma-antitone} gives
\(\Gamma_{W'}(Z)\subseteq\Gamma_{W'}(Y')\). By the law of aggregate demand for
firms and stability of \(Y'\),
\[
|C_F(\Gamma_{W'}(Z))|
\le
|C_F(\Gamma_{W'}(Y'))|
=
|Y'|.
\]
On the worker side, since \(Y'\subseteq Y\cup Y'\) and \(C_{W'}(Y')=Y'\), the
law of aggregate demand gives \(|Y'|\le |C_{W'}(Y\cup Y')|=|Z|\). Hence,
\(|C_F(\Gamma_{W'}(Z))|\le |Z|\). Since \(Z\in Q(W',F)\),
\(Z\subseteq C_F(\Gamma_{W'}(Z))\), and therefore
\(Z=C_F(\Gamma_{W'}(Z))\). Proposition~\ref{prop:stability} implies
\(Z\in S(W',F)\).
\end{proof}

\begin{proof}[Proof of Theorem~\ref{thm:lad-sublattice}]
Let \(Y,Y'\in S(W',F)\). By Theorem~\ref{thm:stable-sublattice-operations},
\[
Y\tilde{\vee}_{W'}Y'=T^*_{W',F}(Y\vee_{W'}Y')
\quad\text{and}\quad
Y\tilde{\wedge}_{W'}Y'=Y\wedge_{W'}Y'.
\]
Under the law of aggregate demand, Proposition~\ref{prop:lad-join-stability}
implies that \(Y\vee_{W'}Y'\in S(W',F)\). Hence,
\[
T^*_{W',F}(Y\vee_{W'}Y')=Y\vee_{W'}Y',
\]
and therefore,
\[
Y\tilde{\vee}_{W'}Y'=Y\vee_{W'}Y'.
\]
The meet equality is already given by Theorem~\ref{thm:stable-sublattice-operations}.
\end{proof}

\begin{proof}[Proof of Corollary~\ref{cor:entrant-firms}]
By Theorem~\ref{thm:explicit-formula},
\(\Phi(Y)=\Pi(Y)\vee_{W'}Y^F\). Hence,
\(\Phi(Y)\succeq_{W'}^B Y^F\). By the fixed-market opposition-of-interests
property, \(Y^F\succeq_F^B\Phi(Y)\).

The worker-side comparison and the law of aggregate demand imply
\(|\Phi(Y)|\ge |Y^F|\), while the firm-side comparison and the law of aggregate demand imply
\(|\Phi(Y)|\le |Y^F|\). Hence, \(|\Phi(Y)|=|Y^F|\). Moreover,
\(Y^F\succeq_F^B\Phi(Y)\) implies, firm by firm, that
\[
|\Phi(Y)_f|\le |Y^F_f|\qquad\text{for every }f\in F.
\]
Together with \(|\Phi(Y)|=|Y^F|\), this yields
\[
|\Phi(Y)_f|=|Y^F_f|\qquad\text{for every }f\in F.
\]

Now let \(f\in F^E\). Since no pre-shock allocation contains contracts
involving \(f\), \(\Pi(Y)_f=\emptyset\). Also
\[
\Phi(Y)=C_{W'}(\Pi(Y)\cup Y^F)\subseteq \Pi(Y)\cup Y^F,
\]
and therefore \(\Phi(Y)_f\subseteq Y^F_f\). The cardinality equality above
implies \(\Phi(Y)_f=Y^F_f\).
\end{proof}
 
\bibliographystyle{econ}
\bibliography{fqs}
\end{document}